\documentclass[prd,aps,nofootinbib,onecolumn,showpacs,12pt]{revtex4-1}
\usepackage{graphicx,epsfig}

\usepackage{epsf}
\usepackage{graphicx}

\begin{document}

\title{ \bf Strong coupling constants of bottom and charmed mesons with scalar, pseudoscalar and axial vector kaons}
\author{ H. Sundu$^{*1}$, J. Y. S\"ung\"u$^{*2}$, S. \c{S}ahin$^{*3}$, N. Yinelek$^{*4}$, K. Azizi$^{\dag5}$\\
 $^{*}$Department of Physics, Kocaeli University, 41380 Izmit,
Turkey\\
$^{\dag}$Physics Division,  Faculty of Arts and Sciences,
Do\u gu\c s University,
 Ac{\i}badem-Kad{\i}k\"oy, \\ 34722 Istanbul, Turkey\\
$^1$e-mail:hayriye.sundu@kocaeli.edu.tr\\
$^2$e-mail:jyilmazkaya@kocaeli.edu.tr\\
$^3$e-mail:095131004@kocaeli.edu.tr\\
$^4$e-mail:neseyinelek@gmail.com\\
$^5$e-mail:kazizi@dogus.edu.tr}

\begin{abstract}
The strong coupling constants,  $g_{D_{s}DK_0^*}$, $g_{B_{s}BK_0^*}$,   $g_{D^{\ast}_{s}D K}$, $g_{B^{\ast}_{s}BK}$, $g_{D^{\ast}_{s}D K_1}$ and $g_{B^{\ast}_{s}BK_1}$, where $K_0^*$, $K$ and $K_1$ are scalar, 
pseudoscalar and axial vector kaon mesons, respectively   are calculated in the framework of three-point QCD sum rules. 
In particular,  the
correlation functions of the considered vertices when  both $B(D)$ and $K_0^*(K)(K_1)$
mesons are off-shell are evaluated. In the case of $K_1$, which is either $K_1(1270)$ or $K_1(1400)$, the mixing between these two states are also taken into account. 
A comparison of the obtained result with the existing prediction on $g_{D^{\ast}_{s}D K}$  as the only coupling constant among the considered vertices, previously calculated in the literature, is also made.
\end{abstract}

\maketitle

%%%%%%%%%%%%%%%%%%%%%%%%%%%%%%%%%%%%%%%%%%%%%%%%%%%%%%%%%%%%%%%%%%%%
%\section{Introduction}
%%%%%%%%%%%%%%%%%%%%%%%%%%%%%%%%%%%%%%%%%%%%%%%%%%%%%%%%%%%%%%%%%%%%
\section{Introduction}

The strong coupling constants among the bottom and charmed  mesons with light scalar, pseudoscalar and axial strange mesons are the main
ingredients in analysis of  their strong interactions.  More accurate determination of these coupling constants is
needed to better understand the  strong interactions among the participated mesons,  construct the strong  potentials among them and obtain knowledge about the nature and structure of the encountered particles.
 Experimentally, it is believed that in the production of the charmonium states like $J/\psi$ and $\psi'$ 
from the $B_c$ or newly discovered charmonium $X$, $Y$ and $Z$ states by the BABAR and BELLE collaborations, there are intermediate  two body states containing  $D$, $D_s$, $D^*$ and
 $D^*_s$ mesons (for example, the kaon can annihilate the charmonium in a nuclear medium to give $D$ and $D_s$ mesons), which decay to the final
$J/\psi$ and $\psi'$ states exchanging
 one or more virtual mesons. A similar story would happen in decays of heavy bottonium. To exactly follow and analyze the procedure in the experiment, we need  to have knowledge about the coupling constants among
the  particles involved.

The strong coupling constants among mesons take place in low energies  very far from the perturbative region, where the strong coupling constant between quarks and gluons is 
large and perturbation theory fails.  Hence in the hadronic scale, one should consult 
to some  nonperturbative methods in QCD to describe nonperturbative phenomena. Among the nonperturbative methods, the QCD sum rules approach \cite{Shifman,ek1,Reinders,Narison} is one of the most powerful, applicable
and attractive one as it is based on QCD Lagrangian and is free of a model dependent parameter. This approach has rendered
 many successful predictions such as its predictions about the vector
 mesons \cite{ek11,ek111,ek1111,ek11111,ek111111}. 
The three point correlation function  has been widely used to calculate many parameters of hadrons (see 
for instance \cite{ek2,ek3,ek4,ek5}). The QCD sum rules for some  strong coupling constants were derived by means of the three point functions in \cite{ek6}. In the present work, we investigate various strong coupling constants 
among bottom (charmed)--bottom strange (charmed strange) mesons with scalar, pseudoscalar and axial vector kaons. Calculation of such coupling constants can help us in understanding the nature of the strong 
interaction among the participating particles.

 In the case of the scalar kaon, we consider the $B_{s}-B-K_0^*$ and $D_{s}-D-K_0^*$ vertices for
both $K_0^*(800)$ and $K_0^*(1430)$. Understanding the internal structure of the scalar mesons
has been a striking issue in the last 30-40 years. Despite their investigation  both theoretically
and experimentally, most  of their properties  are not very clear yet. Detection and identification of the scalar mesons are difficult,
experimentally, so the theoretical and phenomenological works can play a crucial role in this regard. In this work, we also calculate the  coupling constants $g_{B^{\ast}_{s}BK}$ and $g_{D^{\ast}_{s}D K}$ for pseudoscalar $K$.
The next aim in the present work is to consider the vertices $B^*_{s}-B-K_1$ and $D^*_{s}-D-K_1$ for both $K_1(1270)$ and $K_1(1400)$ axial states taking into account their mixture.

Experimentally, the $K_{1} (1270)$ and $K_{1} (1400)$ are the
mixtures of the strange members of two axial-vector SU(3) octets
$^{3}P_{1}(K_{1}^{A})$ and $^{1}P_{1}(K_{1}^{B})$. To avoid any confusion between the $B$ meson and the sign $B$ in the $K_{1}^{B}$, we will use  the $K_{1}^{a(b)}$ instead of $K_{1}^{A(B)}$ in this article. The $K_{1} (1270,
1400)$ are related to the  $K_{1}^{a,b}$ states via
\cite{lee, suzuki}:

\begin{eqnarray}\label{melo}
\mid K_{1}(1270)\rangle&=&\mid K_{1}^{a}\rangle sin \theta+\mid K_{1}^{b}\rangle cos
\theta, \nonumber \\
\mid K_{1}(1400)\rangle&=&\mid K_{1}^{a}\rangle cos \theta-\mid K_{1}^{b}\rangle sin
\theta, \nonumber\\
\end{eqnarray}
where the mixing angle $\theta$ takes the values in the interval
$37^{\circ}\leq\theta\leq58^{\circ}$,
$-58^{\circ}\leq\theta\leq-37^{\circ}$ \cite{lee, suzuki,cenk,
burak, cheng}. The sign ambiguity for the mixing angle is correlated  with
the fact that one can add arbitrary phase to the $\mid K_{1}^{a}\rangle$
and $\mid K_{1}^{b}\rangle$. Studies on $B\rightarrow
K_{1}(1270)\gamma$ and $\tau\rightarrow K_{1}(1270)\nu_{\tau}$ lead to the
following value for $ \theta$ \cite{hatanaka}:
\begin{equation}\label{tetam}
\theta=-(34\pm13)^{\circ}
\end{equation}

In the present work, contributing the quark-quark and quark-gluon condensate diagrams as nonperturbative effects, we evaluate the corresponding correlation functions when  both $B(D)$ and $K_0^*(K)(K_1)$
mesons are off-shell. Note that recently, we have investigated the $D_s^*DK^*(892)$, and
$B_s^*BK^*(892)$ vertices for $K^*$ being the vector meson in the  framework of the three-point QCD sum rules in \cite{K.Azizi}. Moreover, the following coupling constants
have been investigated via  three-point and light cone QCD sum rules   in the literature: $D^*D\pi$
\cite{F.S.Navarra,F.S.Navarra1}, $DD\rho$\cite{M.E.Bracco1},
$DDJ/\psi$ \cite{R.D.Matheus}, $D^*DJ/\psi$  \cite{R.D.Matheus1},
$D^*D^*\pi$ \cite{Z.G.Wang,F.Carvalho}, $D^*D^*J/\psi$
\cite{M.E.Bracco2}, $D_sD^*K, D^*_sDK$ \cite{M.E.Bracco,Z.G.Wang},
$D_0D_sK, D_{s0}DK$\cite{Z.G.Wangg}, $DD\omega$ \cite{L.B.Holanda},
$D^*D^*\rho$ \cite{M.E.Bracco3}, $D^*D\rho$ \cite{B.O.Rodrigues},
$B_{s0}BK, B_{s1}B^*K$ \cite{Z.G.Wang2008}, $ B_{s0}BK$ \cite{nielson.yalvarma},  $a_0\eta\pi^0$, $a_0\eta'\pi^0$
\cite{Z.G.Wang2010},
 $a_0K^+K^-$ \cite{Z.G.Wang2004} and $f_0K^+K^-$ \cite{Z.G.Wang2004, P.Colangelo}.

The outline of the paper is as follows. In section II, introducing responsible correlation functions, we obtain QCD sum rules for the 
strong coupling constant of the considered vertices. For each of the scalar, pseudoscalar and axial kaon cases, we will calculate the  correlation function when both the $B(D)$ and $K_0^*(K)(K_1)$ mesons are 
off-shell. In the case of the $K_1$ meson, first we will calculate the QCD sum rules for the vertices $B^*_{s}-B-K^a_1$ and $D^*_{s}-D-K^b_1$, then using the relations in Eq. (\ref{melo}), we will acquire the QCD
 sum rules for the 
vertices  $B^*_{s}-B-K_1(1270)$ and $D^*_{s}-D-K_1(1400)$. In obtaining the sum rules for physical quantities, we will consider both light quark-light quark and light quark-gluon  condensate diagrams as nonperturbative
contributions. Finally, in section III, we numerically analyze the obtained sum rules for the considered strong coupling constants. We will obtain the numerical values for each coupling constant when both the 
$B(D)$ and $K_0^*(K)(K_1)$ states are off-shell. Then taking the average of the two off-shell cases, we will obtain final numerical values for each coupling constant. In this section, we also compare our result
 on $g_{D^{\ast}_{s}D K}$ with  existing prediction in the literature.

\section{QCD Sum Rules for the Strong Coupling Constants}

In this section, we obtain QCD sum rules for the  strong coupling constants associated with the  $D_{s}(B_{s})-D(B)-K_0^*$,  $D^*_{s}(B^*_{s})-D(B)-K$ and $D^*_{s}(B^*_{s})-D(B)-K_1$ vertices.
 We start our discussion
considering the sufficient correlation functions responsible for the corresponding strong transition involving each $K_0^*$, $K$ and $K_1$ mesons when both $D(B)$ and $K_0^*(K)(K_1)$ are off-shell. The following 
three-point correlation functions describe the considered strong transitions:
\begin{itemize}
 \item correlation functions corresponding to the  $D_{s}(B_{s})-D(B)-K_0^*$ vertex:
\begin{itemize}
\item for  $D(B)$ off-shell:
\begin{eqnarray}\label{CorrelationFunc1}
\Pi^{D (B)}=i^2 \int d^4x~d^4y~ e^{ip^{\prime}\cdot
x}~ e^{iq\cdot y}{\langle}0| {\cal T}\left ( \eta^{K^{\ast}_0}(x)~
\eta^{D(B)}(y)~ \eta^{D_{s}(B_{s})\dag}(0)\right)|0{\rangle},
\end{eqnarray}
\item for  $K^{\ast}_0$ off-shell:
\begin{eqnarray}\label{CorrelationFunc2}
\Pi^{K^{\ast}_0}=i^2 \int
d^4x~d^4y~ e^{ip^{\prime}\cdot x}~ e^{iq\cdot y}{\langle}0| {\cal
T}\left ( \eta^{D(B)}(x)~ \eta^{K^{\ast}_0}(y)~
\eta^{D_{s}(B_{s})\dag}(0)\right)|0{\rangle},
\end{eqnarray}

\end{itemize}
\item correlation functions corresponding to the  $D^*_{s}(B^*_{s})-D(B)-K$ vertex:
\begin{itemize}
\item for  $D(B)$ off-shell:
\begin{eqnarray}\label{CorrelationFunc1}
\Pi_{\mu}^{D (B)}=i^2 \int d^4x~d^4y~
e^{ip^{\prime}\cdot x}~ e^{iq\cdot y}{\langle}0| {\cal T}\left (
\eta^{K}(x)~ \eta^{D(B)}(y)~
\eta^{D_{s}^{\ast}(B_{s}^{\ast})\dag}_{\mu}(0)\right)|0{\rangle},
\end{eqnarray}
\item for  $K$ off-shell:
\begin{eqnarray}\label{CorrelationFunc2}
\Pi_{\mu}^{K}=i^2 \int d^4x~d^4y~ e^{ip^{\prime}\cdot
x}~ e^{iq\cdot y}{\langle}0| {\cal T}\left ( \eta^{D(B)}(x)~
\eta^{K}(y)~
\eta^{D_{s}^{\ast}(B_{s}^{\ast})\dag}_{\mu}(0)\right)|0{\rangle},
\end{eqnarray}
\end{itemize}
\item correlation functions corresponding to the $D^*_{s}(B^*_{s})-D(B)-K_1$ vertex:
\begin{itemize}
\item for  $D(B)$ off-shell:
\begin{eqnarray}\label{CorrelationFunc1}
\Pi_{\mu\nu}^{D (B)}=i^2 \int d^4x~d^4y~
e^{ip^{\prime}\cdot x}~ e^{iq\cdot y}{\langle}0| {\cal T}\left (
\eta^{K_1}_{\nu}(x)~ \eta^{D(B)}(y)~
\eta^{D_{s}^{\ast}(B_{s}^{\ast})\dag}_{\mu}(0)\right)|0{\rangle},\nonumber\\
\end{eqnarray}
\begin{eqnarray}\label{CorrelationFunc1}
\Pi_{\mu\nu\nu'}^{D (B)}=i^2 \int d^4x~d^4y~
e^{ip^{\prime}\cdot x}~ e^{iq\cdot y}{\langle}0| {\cal T}\left (
\eta^{K_1}_{\nu\nu'}(x)~ \eta^{D(B)}(y)~
\eta^{D_{s}^{\ast}(B_{s}^{\ast})\dag}_{\mu}(0)\right)|0{\rangle},\nonumber\\
\end{eqnarray}
\item for  $K_1$ off-shell:
\begin{eqnarray}\label{CorrelationFunc2}
\Pi_{\mu\nu}^{K_1}=i^2 \int d^4x~d^4y~ e^{ip^{\prime}\cdot
x}~ e^{iq\cdot y}{\langle}0| {\cal T}\left ( \eta^{D(B)}(x)~
\eta^{K_1}_{\nu}(y)~
\eta^{D_{s}^{\ast}(B_{s}^{\ast})\dag}_{\mu}(0)\right)|0{\rangle},\nonumber\\
\end{eqnarray}
\begin{eqnarray}\label{CorrelationFunc2}
\Pi_{\mu\nu\nu'}^{K_1}=i^2 \int d^4x~d^4y~ e^{ip^{\prime}\cdot
x}~ e^{iq\cdot y}{\langle}0| {\cal T}\left ( \eta^{D(B)}(x)~
\eta^{K_1}_{\nu\nu'}(y)~
\eta^{D_{s}^{\ast}(B_{s}^{\ast})\dag}_{\mu}(0)\right)|0{\rangle},\nonumber\\
\end{eqnarray}
\end{itemize}
\end{itemize}
where ${\cal T}$ is  the time ordering
product, $q$ is the momentum of the off-shell state, $p'$ is the momentum of the final on-shell state. We will set the momentum of the initial state as $p=p'+q$. In the vertex containing   $K_1$ meson, 
we have  two correlation
functions for both off-shell cases since this meson couples into two interpolating currents $\eta^{K_1}_{\nu}$ and $\eta^{K_1}_{\nu\nu'}$. We will define these couplings in terms of 
G-parity conserving and G-parity violating
 decay constants later.
The  interpolating currents, which produce the considered mesons from the vacuum with the same quantum numbers as the interpolating currents can be written in terms of
the quark field operators as following form:
\begin{eqnarray}\label{mesonintfield}
\eta^{K^*_0}(x)&=& \overline{s}(x)Uu(x),\nonumber \\
\eta^{K}(x)&=& \overline{s}(x)\gamma_{5}u(x),
\nonumber \\
\eta^{K_1}_\nu(x)&=& \overline{s}(x)\gamma_\nu\gamma_{5}u(x),
\nonumber \\
\eta^{K_1}_{\nu\nu'}(x)&=& \overline{s}(x)\sigma_{\nu\nu'}\gamma_{5}u(x),
\nonumber \\
\eta^{D(B)}(x)&=& \overline{u}(x)\gamma_{5}c(b)(x),
\nonumber \\
\eta^{D_{s}(B_{s})}(x)&=&
\overline{s}(x)\gamma_{5}c(b)(x),\nonumber \\
\eta^{D_{s}^{\ast}(B_{s}^{\ast})}_{\mu}(x)&=&
\overline{s}(x)\gamma_{\mu}c(b)(x),
\end{eqnarray}
where $U$ stands for unit matrix and u, s, c and b are the up, strange, charm and bottom quark
fields, respectively. 

According to general philosophy of the QCD sum rules, we  calculate the aforementioned correlation functions in two different representations. In physical or phenomenological representation, 
we calculate them  in terms of hadronic parameters. In QCD or theoretical representation, we evaluate them  in terms of  QCD degrees of freedom like quark masses, quark condensates, etc.  with the help of
 the operator
product expansion (OPE), where the perturbative and nonperturbative contributions are separated. The QCD sum rules for strong coupling constants are obtained  equating these two different representations through
dispersion relation. To suppress  contributions of the higher states and continuum, we will apply double Borel transformation with respect to the momentum squared of the initial
 and final on-shell states to both sides of the obtained sum rules.

First, let us focus on the calculation of the physical sides of the aforesaid
correlation functions for example when D(B) meson is off-shell. Saturating
the correlation functions with the complete sets of three participating particles and isolating the ground states and after some straightforward calculations, we obtain:
\begin{itemize}
 \item physical representation corresponding to the $D_{s}(B_{s})-D(B)-K_0^*$ vertex:
\begin{eqnarray}\label{CorrelationFuncPhys1}
\Pi^{D (B)}&=&\frac{{\langle}0|\eta^{K_0^*}|K_0^*
(p^{\prime}){\rangle} {\langle}0|\eta^{D(B)}|D(B) (q){\rangle}
{\langle}K_0^*(p^{\prime}) D(B)
(q)|D_{s}(B_{s})(p){\rangle}
{\langle}D_s(B_{s})
(p)|\eta^{D_s(B_{s})}|0{\rangle}}
{(q^2-m_{D(B)}^2)
(p^2-m_{D_s(B_{s})}^2)({p^{\prime}}^{2}-m_{K_0^*}^2)}
\nonumber \\ &+&...,
\end{eqnarray}
\item physical representation corresponding to the $D^*_{s}(B^*_{s})-D(B)-K$ vertex:
\begin{eqnarray}\label{CorrelationFuncPhys11}
\Pi_{\mu}^{D (B)}&=&\frac{{\langle}0|\eta^{K}|K
(p^{\prime}){\rangle} {\langle}0|\eta^{D(B)}|D(B) (q){\rangle}
{\langle}K(p^{\prime}) D(B)
(q)|D_{s}^{\ast}(B_{s}^{\ast})(p,\epsilon){\rangle}
{\langle}D_s^{\ast}(B_{s}^{\ast})
(p,\epsilon)|\eta^{D_s^{\ast}(B_{s}^{\ast})}_{\mu}|0{\rangle}}
{(q^2-m_{D(B)}^2)
(p^2-m_{D_s^{\ast}(B_{s}^{\ast})}^2)({p^{\prime}}^{2}-m_{K}^2)}
\nonumber \\ &+&...,
\end{eqnarray}
\item physical representation corresponding to the $D^*_{s}(B^*_{s})-D(B)-K_1^{a(b)}$ vertex:
\begin{eqnarray}\label{CorrelationFuncPhys11}
\Pi_{\mu\nu}^{D (B)}&=&\frac{{\langle}0|\eta^{D(B)}|D(B) (q){\rangle}{\langle}D_s^{\ast}(B_{s}^{\ast})
(p,\epsilon)|\eta^{D_s^{\ast}(B_{s}^{\ast})}_{\mu}|0{\rangle}}{(q^2-m_{D(B)}^2)
(p^2-m_{D_s^{\ast}(B_{s}^{\ast})}^2)}\nonumber\\&\times&\Bigg[\frac{{\langle}0|\eta^{K_1}_\nu|K_1^{a}
(p^{\prime},\epsilon'){\rangle}{\langle}K_1^{a}(p^{\prime},\epsilon') D(B)
(q)|D_{s}^{\ast}(B_{s}^{\ast})(p,\epsilon){\rangle}}{({p^{\prime}}^{2}-m_{K_1^{a}}^2)}\nonumber\\&+&\frac{{\langle}0|\eta^{K_1}_\nu|K_1^{b}
(p^{\prime},\epsilon'){\rangle}{\langle}K_1^{b}(p^{\prime},\epsilon') D(B)
(q)|D_{s}^{\ast}(B_{s}^{\ast})(p,\epsilon){\rangle}}{({p^{\prime}}^{2}-m_{K_1^{b}}^2)}\Bigg]
+...,
\end{eqnarray}
\begin{eqnarray}\label{CorrelationFuncPhys11}
\Pi_{\mu\nu\nu'}^{D (B)}&=&\frac{{\langle}0|\eta^{D(B)}|D(B) (q){\rangle}{\langle}D_s^{\ast}(B_{s}^{\ast})
(p,\epsilon)|\eta^{D_s^{\ast}(B_{s}^{\ast})}_{\mu}|0{\rangle}}{(q^2-m_{D(B)}^2)
(p^2-m_{D_s^{\ast}(B_{s}^{\ast})}^2)}\nonumber\\&\times&\Bigg[\frac{{\langle}0|\eta^{K_1}_{\nu\nu'}|K_1^{a}
(p^{\prime},\epsilon'){\rangle}{\langle}K_1^{a}(p^{\prime},\epsilon') D(B)
(q)|D_{s}^{\ast}(B_{s}^{\ast})(p,\epsilon){\rangle}}{({p^{\prime}}^{2}-m_{K_1^{a}}^2)}\nonumber\\&+&\frac{{\langle}0|\eta^{K_1}_{\nu\nu'}|K_1^{b}
(p^{\prime},\epsilon'){\rangle}{\langle}K_1^{b}(p^{\prime},\epsilon') D(B)
(q)|D_{s}^{\ast}(B_{s}^{\ast})(p,\epsilon){\rangle}}{({p^{\prime}}^{2}-m_{K_1^{b}}^2)}\Bigg]
+...,
\end{eqnarray}

\end{itemize}
where .... represents the contributions of the higher states and 
continuum, and $\epsilon$ and $\epsilon'$  are the polarization vectors associated with the
$D_{s}^{\ast}$ and $K_1^{a(b)}$ mesons, respectively. From the above equations it is clear that to proceed we need to define the following matrix elements in terms of decay constants as well as
 strong coupling constants:
 
\begin{eqnarray}\label{transitionamp}
{\langle}
0|\eta^{K^{\ast}_0}|K^{\ast}_0(p^{\prime}){\rangle}&=&m_{K^{\ast}_0}
f_{K^{\ast}_0},
\nonumber\\
{\langle} 0|\eta^{K}|K(p^{\prime}){\rangle}&=&\frac{m_K^2
f_K}{m_u+m_s},
\nonumber\\
{\langle}0|\eta^{D(B)}|D(B)(q){\rangle}&=&\frac{m_{D(B)}^2~f_{D(B)}}{m_{c(b)}+m_u},
\nonumber\\
{\langle}0|\eta^{D_s(B_s)}|D_s(B_s)(q){\rangle}&=&\frac{m_{D_s(B_s)}^2~f_{D_s(B_s)}}{m_{c(b)}+m_s},
\nonumber\\
{\langle}D_{s}^{\ast}(B_{s}^{\ast})(p,\epsilon)
 |\eta^{D_{s}^{\ast}(B_{s}^{\ast})}_{\mu}|0{\rangle}&=&m_{D_{s}^{\ast}
 (B_{s}^{\ast})}
 f_{D_{s}^{\ast}(B_{s}^{\ast})}{\epsilon^{\ast}}_{\mu},
 \nonumber\\
{\langle}K^{\ast}_0(p^{\prime})D(B)(q)|D_{s}(B_{s})(p){\rangle}&=&
g_{D_{s}D K^{\ast}_0(B_{s}B K^{\ast}_0)}p\cdot p^{\prime},\nonumber\\
{\langle}K(p^{\prime})D(B)(q)|D^{\ast}_{s}(B^{\ast}_{s})(p,\epsilon){\rangle}&=&g_{D^{\ast}_{s}D
K(B^{\ast}_{s}B K)} (p^{\prime}-q)\cdot \epsilon,\nonumber\\
{\langle}K_1^{a(b)}(p^{\prime},\epsilon') D(B)
(q)|D_{s}^{\ast}(B_{s}^{\ast})(p,\epsilon){\rangle}&=&g_{D^{\ast}_{s}D
K_1^{a(b)}(B^{\ast}_{s}B K_1^{a(b)})}\{m^2_{D_s^*(B_s^*)}(\epsilon^{'*}.\epsilon)+(\epsilon^{'*}.p)(\epsilon.p')\},\nonumber\\
\end{eqnarray}
and 
\begin{eqnarray}
 \left. \begin{array}{l}
{\langle}
0|\eta^{K_1}_\nu|K^{a}_1(p^{\prime},\epsilon'){\rangle}=m_{K^{a}_1}
f_{K^{a}_1}\epsilon'_\nu,
\nonumber\\
{\langle}
0|\eta^{K_1}_{\nu\nu'}|K^{b}_1(p^{\prime},\epsilon'){\rangle}=f_{K_{1}^{b^{\perp}}}(\epsilon'_{\nu}p'_{\nu'}-\epsilon'_{\nu'}p'_{\nu})\end{array} \right\}&&\mbox{ G-parity conserving definitions,}
\nonumber\\
\left. \begin{array}{l}
{\langle}
0|\eta^{K_1}_{\nu\nu'}|K^{a}_1(p^{\prime},\epsilon'){\rangle}=f_{K_{1}^{a}}~a_{0}^{\perp,K_{1}^{a}}(\epsilon'_{\nu}p'_{\nu'}-\epsilon'_{\nu'}p'_{\nu}),
\nonumber\\
{\langle}
0|\eta^{K_1}_\nu|K^{b}_1(p^{\prime},\epsilon'){\rangle}=m_{K^b_1}
f_{K^{b\bot}_1}a_0^{\parallel,K^{b}_1}\epsilon'_\nu\end{array} \right\}&&\mbox{ G-parity violating definitions,}
\nonumber\\
\end{eqnarray}
where  $f_{K_0^*}$, $f_K$,
$f_{D(B)}$, $f_{D_s(B_s)}$ and $f_{D_s^*(B_s^*)}$ are leptonic decay constants of the $K_0^*$, $K$, $D(B)$, $D_s(B_s)$ and $D_s^*(B_s^*)$ mesons, respectively. The $f_{K^{a}_1}$  and $f_{K^{b\bot}_1}$
are decay constants encountered to the calculations from both definitions of the G-parity conserving and violating matrix elements for the axial $K^{a}_1$ and $K^{b}_1$  states 
(for more details see \cite{busey,lee,hatanaka,azizi2}). 
 The $a_{0}^{\perp,K_{1}^{a}}$ and $a_0^{\parallel,K^{b}_1}$ are   zeroth order Gegenbauer moments. In the above equations, the 
$g_{D_{s}D K^{\ast}_0(B_{s}B K^{\ast}_0)}$,
$g_{D^{\ast}_{s}D
K(B^{\ast}_{s}B K)}$ and $g_{D^{\ast}_{s}D
K_1^{a(b)}(B^{\ast}_{s}B K_1^{a(b)})}$ are strong coupling constants, which we are going to obtain  QCD sum rules for them in this section.

Using
Eqs. (\ref{CorrelationFuncPhys1})-(\ref{transitionamp}), and summing over the polarization vectors using the
\begin{eqnarray}\label{polvec}
\epsilon'_\nu\epsilon^{'*}_\theta&=&-g_{\nu\theta}+\frac{p'_{\nu}p'_{\theta}}{m_{K^{a(b)}_{1}}^2},\nonumber\\
\epsilon_\eta\epsilon^{*}_\mu&=&-g_{\eta\mu}+\frac{p_{\eta}p_{\mu}}{m_{D_{s}^{\ast}(B_{s}^{\ast})}^2},
\end{eqnarray}
the final  physical
representations of the correlation functions for each vertices in the case of  D(B) off-shell  is obtained  as:
\begin{itemize}
 \item  $D_{s}(B_{s})-D(B)-K_0^*$ vertex:
\begin{eqnarray}\label{CorrelationFuncPhys2}
\Pi^{D (B)}&=&g^{D(B)}_{D_{s}D K^{\ast}_0(B_{s}B
K^{\ast}_0)}(q^2)\frac{
f_{K^{\ast}_0}m_{K^{\ast}_0}\frac{f_{D(B)}m_{D(B)}^2}{m_{c(b)}+m_u}\frac{f_{D_{s}(B_{s})}m_{D_{s}(B_{s})}^2}
{m_{c(b)}+m_s} }{2~(q^2-m_{D(B)}^2)({p^{\prime}}^{2}-m_{K^{\ast}_0}^2)
(p^2-m_{D_s(B_{s})}^2)}(m_{D_{s}(B_{s})}^2
+m_{K^{\ast}_0}^2-q^2)\nonumber\\&+& ....,
\end{eqnarray}
\item  $D^*_{s}(B^*_{s})-D(B)-K$ vertex:
\begin{eqnarray}\label{CorrelationFuncPhys2}
\Pi_{\mu}^{D (B)}&=&g^{D(B)}_{D^{\ast}_{s}D
K(B^{\ast}_{s}B K)}(q^2)\frac{f_{D^{\ast}_{s}(B^{\ast}_{s})}m_{D^{\ast}_{s}(B^{\ast}_{s})}
 \frac{f_{K}m_{K}^2}{m_s+m_u} \frac{f_{D(B)}m_{D(B)}^2}{m_{c(b)}+m_u}
}{(q^2-m_{D(B)}^2)({p^{\prime}}^{2}-m_{K}^2)
(p^2-m_{D_s^{\ast}(B_{s}^{\ast})}^2)
}
\Big[\Big(1+\frac{m_{K}^2-q^2}{m^2_{D^{\ast}_{s}}}\Big)
p_{\mu}-2p_{\mu}^{\prime}\Big]\nonumber\\&+& ....,
\end{eqnarray}
\item  $D^*_{s}(B^*_{s})-D(B)-K_1$ vertex:
\begin{eqnarray}
\Pi_{\mu\nu}^{D (B)}&=&\Bigg(g^{D(B)}_{D^{\ast}_{s}D
K_1^{a}(B^{\ast}_{s}B K_1^{a})}(q^2)\frac{m_{K^{a}_1}
f_{K^{a}_1}}{({p^{\prime}}^{2}-m_{K_1^{a}}^2)}+g^{D(B)}_{D^{\ast}_{s}D
K_1^{b}(B^{\ast}_{s}B K_1^{b})}(q^2)\frac{m_{K^b_1}
f_{K^{b\bot}_1}a_0^{\parallel,K^{b}_1}}{({p^{\prime}}^{2}-m_{K_1^{b}}^2)}\Bigg)\nonumber\\&\times&\frac{\frac{f_{D(B)}m_{D(B)}^2}{m_{c(b)}+m_u}f_{D^{\ast}_{s}(B^{\ast}_{s})}m_{D^{\ast}_{s}(B^{\ast}_{s})}}{(q^2-m_{D(B)}^2)
(p^2-m_{D_s^{\ast}(B_{s}^{\ast})}^2)}\Bigg\{m^2_{D_s^*(B_s^*)}g_{\mu\nu}+ \mbox{\rm other structures}\Bigg\}+...,
 \end{eqnarray}
\begin{eqnarray}
\Pi_{\mu\nu\nu'}^{D (B)}&=&\Bigg(g^{D(B)}_{D^{\ast}_{s}D
K_1^{a}(B^{\ast}_{s}B K_1^{a})}(q^2)\frac{f_{K_{1}^{a}}~a_{0}^{\perp,K_{1}^{a}}}{({p^{\prime}}^{2}-m_{K_1^{a}}^2)}+g^{D(B)}_{D^{\ast}_{s}D
K_1^{b}(B^{\ast}_{s}B K_1^{b})}(q^2)\frac{f_{K_{1}^{b^{\perp}}}}{({p^{\prime}}^{2}-m_{K_1^{b}}^2)}\Bigg)\nonumber\\&\times&\frac{\frac{f_{D(B)}m_{D(B)}^2}{m_{c(b)}+m_u}f_{D^{\ast}_{s}(B^{\ast}_{s})}m_{D^{\ast}_{s}(B^{\ast}_{s})}}{(q^2-m_{D(B)}^2)
(p^2-m_{D_s^{\ast}(B_{s}^{\ast})}^2)}\Bigg\{m^2_{D_s^*(B_s^*)}(g_{\mu\nu}p'_{\nu'}-g_{\mu\nu'}p'_{\nu})+ \mbox{\rm other structures}\Bigg\}+...,\nonumber\\
 \end{eqnarray}
\end{itemize}
where to calculate the coupling constants, we will choose the structures
$p_{\mu}$ and $g_{\mu\nu}(g_{\mu\nu}p'_{\nu'}-g_{\mu\nu'}p'_{\nu})$ from both sides of the correlation functions corresponding to the vertices containing the $K$ and $K_1$ with current $\eta_\nu^{K_1}$ 
($K_1$ with current $\eta_{\nu\nu'}^{K_1}$), respectively. From the similar way, one can easily obtain
 the physical representations of the correlation functions associated with the $K_0^*(K)(K_1)$ off-shell.

\begin{figure}[h!]
\begin{center}
\includegraphics[width=15cm]{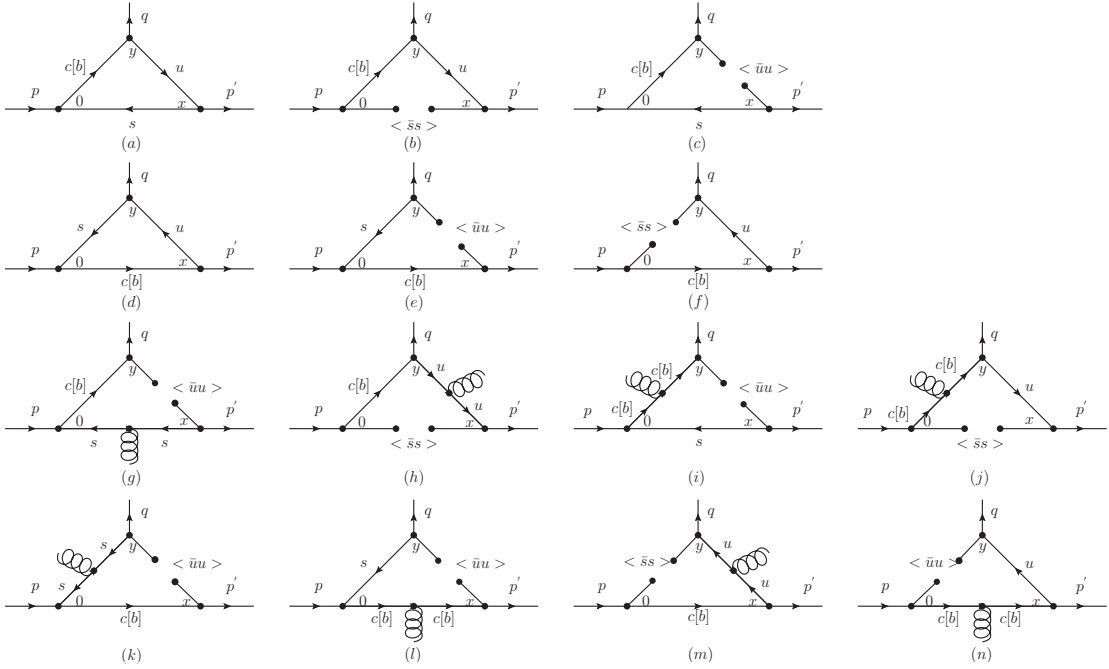}
\end{center}
\caption{Diagrams considered in the calculations. The first and third line diagrams refer to the $B(D)$ off-shell and the second and fourth line diagrams show the case when $K_0^*(K)(K_1)$ is  off-shell.} \label{Figure}
\end{figure}

Now, we concentrate to calculate the QCD or theoretical  sides of the considered correlation functions. The QCD representations of  the correlation functions are 
obtained in deep Euclidean region, where $p^2\rightarrow-\infty$
and ${p^{\prime}}^2\rightarrow-\infty$ via OPE. For this aim, each correlation function in QCD side is written in terms of the perturbative  and non-perturbative parts as:
\begin{eqnarray}\label{CorrelationFuncQCD}
\Pi^{i}&=& \Pi^i_{per}+\Pi^i_{nonper}
\end{eqnarray}
 where $i$ stands for $D(B)$ or $K_0^*(K)(K_1)$ and the perturbative parts are defined in terms of double dispersion
integral as following:
\begin{eqnarray}\label{CorrelationFuncQCDPert}
\Pi^i_{per}&=&-\frac{1}{4 \pi^{2}} \int ds^{\prime} \int ds
\frac{\rho^i(s,s^{\prime}, q^2)}{(s-p^2)
(s^{\prime}-{p^{\prime}}^2)}+\mbox{subtraction terms},
\end{eqnarray}
where $\rho^i(s,s^{\prime}, q^2) $ are called spectral densities. In
order to obtain the spectral density, we need to calculate the bare
loop diagrams (a) and (d) in Fig. (\ref{Figure}) for $D(B)$ and $K_0^*(K)(K_1)$
off-shell, respectively. We calculate these diagrams in terms of the
usual Feynman integrals by the help of the Cutkosky
rules, where  the quark propagators are replaced by Dirac delta
function, i.e.,  $\frac{1}{q^2-m^2}\rightarrow (- 2\pi i) \delta(q^2-m^2)$.
As a result, the spectral
densities are obtained as follows:
\begin{itemize}
 \item  $D_{s}(B_{s})-D(B)-K_0^*$ vertex:
\begin{itemize}
\item $D(B)$ off-shell:
\begin{eqnarray}\label{SpecDenstDB}
\rho^{D(B)}(s,s^{\prime},q^2)&=&\frac{N_c}{2~\lambda^{1/2}(s,s^{\prime},q^2)}
\left\{m_s\left(m_u(m_s+m_u)-q^2\right)-sm_u
\right.\nonumber\\&-&\left.m_{c(b)}\left((m_s+m_u)^2-s^{\prime}-m_{c(b)}(m_s+m_u)\right)\right\},
\end{eqnarray}
\item $K_0^*$ off-shell:
\begin{eqnarray}\label{SpecDenstK}
\rho^{K^*_0}(s,s^{\prime},q^2)&=&\frac{N_c}{2~\lambda^{1/2}(s,s^{\prime},q^2)}
\left\{m_ss^{\prime}+m_{c(b)}\left((m_s+m_u)^2-q^2\right)\right.\nonumber\\&-&\left.m_{c(b)}^2
(m_s+m_u)+m_u\left(s-m_s(m_s+m_u)\right)\right\},
\end{eqnarray}
\end{itemize}
\item  $D^*_{s}(B^*_{s})-D(B)-K$ vertex:
\begin{itemize}
\item $D(B)$ off-shell:
\begin{eqnarray}\label{SpecDenstD}
\rho^{D(B)}(s,s^{\prime},q^2)&=&\frac{N_c}{\lambda^{3/2}(s,s^{\prime},q^2)}
\Big[(m_u-m_s)(q^2-s)\Big(m_{c(b)}
m_s^2+m_u\Big(s-m_s^2-q^2\Big)\Big)\nonumber\\&-&s^{\prime}
\Big(-m_s^3 m_u
+2m_{c(b)}^3(m_u-m_s)-2m_s^2q^2+m_{c(b)}^2(2 m_s
m_u+q^2-s)\nonumber\\
&+&q^2(s-q^2)+m_s m_u(s+q^2)
+m_{c(b)}
(m_s-m_u)(m_s^2+q^2+s)\Big)\nonumber\\
&-&s^{\prime^2}(m_{c(b)}^2-m_{c(b)}
m_s+m_{c(b)} m_u+q^2)\Big],
\end{eqnarray}
\item $K$ off-shell:
\begin{eqnarray}\label{SpecDenstDKs}
\rho^{K}(s,s^{\prime},q^2)&=&\frac{N_c}{\lambda^{3/2}
(s,s^{\prime},q^2)}\Big[(m_{c(b)}-m_u)(q^2-s)\Big(m_{c(b)}^2(m_{c(b)}-m_u)+m_u(-m_s
m_u -q^2\nonumber\\
&+&s)\Big)+\Big(m_{c(b)}^3(m_s-m_u)
+2m_s^3m_u+m_{c(b)}^2
(-m_sm_u-2q^2)+m_s^2(q^2-s)\nonumber\\
&+&q^2(s-q^2)-m_s m_u
(q^2+s)
+m_{c(b)}(-2m_s^3+2m_s^2m_u+m_u(q^2+s)\nonumber\\
&+&m_s(q^2+s))\Big)
s^{\prime}+(-m_{c(b)} m_s+m_s^2+m_s m_u+q^2)s^{\prime^2}\Big],
\end{eqnarray}
\end{itemize}
\item  $D^*_{s}(B^*_{s})-D(B)-K_1$ vertex:
\begin{itemize}
\item $D(B)$ off-shell:
\begin{eqnarray}\label{spectraldensity D ve B icin}
\rho_1^{D(B)}(s,s',q^2)&=&-2N_c
I_0(s,s',q^2)\Big[2m_s^3+2m_s^2m_u-2m_{c(b)}m_s\Big(m_s+m_u\Big)
\nonumber\\&+&m_s\Big(s+s'-q^2\Big) +4A\Big(m_{c(b)}-m_u\Big)
\nonumber\\
&+&B\Big(2 m_u s+m_{c(b)}(q^2-s-s')+m_s(-q^2+3s+s')\Big)
\nonumber\\
&+&C\Big(-2 m_{c(b)}s'+m_u(-q^2+s+s')+m_s(-q^2+s+3s')\Big)\Big],\nonumber\\
\end{eqnarray}
\begin{eqnarray}\label{spectraldensity D ve B icin.ikinci akim}
\rho_2^{D(B)}(s,s',q^2)&=&4N_c I_0(s,s',q^2)\Big[2A+(m_{c(b)}
m_s-m_s^2)-(B+2G)s+H(q^2-s-s')
\nonumber\\
&+&C\Big(m_{c(b)} (m_s+m_u)-m_s(m_s+m_u)
+q^2-s-s'\Big)\Big],\nonumber\\
\end{eqnarray}

\item $K_1$ off-shell:
\begin{eqnarray}\label{spectraldensity K1A ve K1B icin}
\rho_1^{K_1}(s,s',q^2)&=&2N_c
I_0(s,s',q^2)\Big[2m_{c(b)}\Big(m_{c(b)}^2+m_sm_u-m_{c(b)}(m_s+m_u)+s'\Big)
\nonumber\\&-&4D(m_{c(b)}-m_{u})
+E(2m_{c(b)}-m_s-m_u)(q^2-s-s')\nonumber\\&+&2Fs'(-2m_{c(b)}+m_s+m_u)\Big],
\end{eqnarray}
\begin{eqnarray}\label{spectraldensity K1A ve K1B icin2}
\rho_2^{K_1}(s,s',q^2)&=&4N_c I_0(s,s',q^2)\Big[-2D+m_{c(b)}^2-m_{c(b)} m_s-Es
\nonumber\\&-& F\Big(m_{c(b)} (m_{c(b)}-m_s-m_u)+m_s m_u\Big)\Big],
\end{eqnarray}
\end{itemize}
\end{itemize}
where $\rho_1$ and  $\rho_2$ correspond to the currents $\eta_\nu^{K_1}$ and $\eta_{\nu\nu'}^{K_1}$, respectively and
\begin{eqnarray}\label{A}\
A&=&\frac{1}{2\Delta}\{m_s^4q^2+m_{c(b)}^4s'+q^2ss'+m_{c(b)}^2\Big(m_s^2(-q^2+s-s')
+s'(-q^2-s+s') \Big)\nonumber\\&-&m_s^2q^2(-q^2+s+s')\},
\nonumber\\
B&=&\frac{1}{\Delta}\{m_s^2\Big(q^2-s+s'\Big)-s'\Big(2m_{c(b)}^2-q^2-s+s'\Big)\},
\nonumber\\
C&=&\frac{1}{\Delta}\{m_{c(b)}^2\Big(-q^2+s+s'\Big)+s\Big(q^2-s+s'\Big)-m_s^2
\Big(-q^2-s+s'\Big)\},\nonumber\\
D&=&\frac{1}{2\Delta}\{m_{c(b)}^4
q^2+m_s^4s'+q^2ss'-m_{c(b)}^2q^2(-q^2+s+s')+m_s^2\Big(m_{c(b)}^2
(-q^2+s-s')\nonumber\\&+&s'(-q^2-s+s')\Big)\},
\nonumber\\
E&=&\frac{1}{\Delta}\{s'(2m_s^2-q^2-s+s')-m_{c(b)}^2(q^2-s+s')\},
\nonumber\\
F&=&\frac{1}{\lambda(s, s',
q^2)}\{(-m_s^2+m_{c(b)}^2+s)(-q^2+s+s')-2s(m_{c(b)}^2+s')\},\nonumber\\
G&=&\frac{1}{\Delta^2}\{3m_{c(b)}^4s'(q^2-s-s')+m_s^4\Big(2q^4-(s-s')^2
-q^2(s+s')\Big)-ss'\Big(-2q^4+(s-s')^2
\nonumber\\
&+&q^2(s+s')\Big)+m_s^2\Big(q^6- q^4(s+s')+(s-s')^2(s+s')-q^2(s^2-6s
s'+s'^2)\Big)-2m_{c(b)}^2
\nonumber\\
&\Big(&s'(q^4-2s^2+q^2(s-2s')+ss'+s'^2)+m_s^2 (q^4+s^2+s
s'-2s'^2+q^2(-2s+s'))\Big)\},
\nonumber\\
H&=&\frac{1}{\Delta^2}\{s^2\Big(q^4-2q^2(s-2s')+(s-s')^2\Big)+m_s^4
\Big(q^4+2q^2(2s-s')+(s-s')^2\Big)+m_{c(b)}^4
\nonumber\\
&\Big(&q^4+s^2+4ss' +s'^2-2q^2(s+s')\Big)-2m_s^2 s
\Big(-2q^4+(s-s')^2+q^2(s+s')\Big)-2m_{c(b)}^2
\nonumber\\
&\Big(&m_s^2(q^4-2s^2+q^2(s-s')
+ss'+s'^2)+s\Big(q^4+s^2+ss'-2s'^2+q^2(-2s+s')\Big)\Big) \},
\end{eqnarray}
and
\begin{eqnarray}\label{I0integrali}\
I_{0}(s, s', q^2)&=&\frac{1}{4\lambda^{1/2}(s, s', q^2)}
\nonumber\\
\Delta&=&q^4+(s-s')^2-2q^2(s+s')
\nonumber\\
\lambda(a,b,c)&=&a^2+b^2+c^2-2ac-2bc-2ab.
\end{eqnarray}
In the spectral densities, $N_c=3$ is the color
number and we have kept terms linear in $m_u$.

Now, we proceed to calculate  the nonperturbative contributions in QCD side.
We consider the quark-quark and quark-gluon condensate diagrams presented as (b), (c),
(e), (f), (g), (h), (i), (j), (k), (l), (m) and (n) in Fig. (\ref{Figure}).   Contributions of the diagrams (c), (e),
(f), (g), (i), (k), (l), (m) and (n) in Fig. (\ref{Figure}) are zero
since applying double Borel transformation with respect to the both
variables $p^2$ and $p^{\prime^2}$ will kill their contributions because of
 only one variable appearing in the denominator in these cases.
Hence, we consider contributions of only diagrams (b), (h) and (j) in Fig.
(\ref{Figure}). As a result, we
obtain:
\begin{itemize}
 \item  $D_{s}(B_{s})-D(B)-K_0^*$ vertex:
\begin{itemize}
\item $D(B)$ off-shell:
\begin{eqnarray}\label{NonpertResult}
\Pi_{nonper}^{D(B)}&=&\frac{\langle
\overline{s}s\rangle}{2}\Big\{\frac{2m_{c(b)}m_u-m_{c(b)}^2+q^2}
{rr'}
-\frac{1}{r}-\frac{1}{r'}+
\frac{m_0^2(4m_{c(b)}m_u-m_{c(b)}^2+q^2)}{4r^2
r'}
 \nonumber \\
&-&\frac{m_0^2}{4rr'}-\frac{m_0^2}
{4r^2}+\frac{m_0^2(m_{c(b)}^2-4m_{c(b)}m_u-q^2)}
{4rr'^2}+\frac{m_0^2}{4r^2}
+\frac{m_0^2}{4rr'}\Big\},
\end{eqnarray}
\item $K_0^*$ off-shell:
\begin{eqnarray}
\Pi_{nonper}^{K_0^*}&=&0,
\end{eqnarray}
\end{itemize}
\item  $D^*_{s}(B^*_{s})-D(B)-K$ vertex:
\begin{itemize}
\item $D(B)$ off-shell:
\begin{eqnarray}\label{CorrelationFuncNonpertD}
\Pi_{nonper}^{D(B)}&=&-{\langle}\overline{s}s{\rangle}\Big\{\frac{m_u}
{rr'}+\frac{m_0^2\,\, m_u}{4\,
r^2\,r'
}+\frac{m_0^2\,\,m_u}{2\,r\,r'^2}\Big\},
\end{eqnarray}
\item $K$ off-shell:
\begin{eqnarray}\label{CorrelationFuncNonpertB}
\Pi_{nonper}^{K}&=&0,
\end{eqnarray}
\end{itemize}
\item  $D^*_{s}(B^*_{s})-D(B)-K_1$ vertex:
\begin{itemize}
\item $D(B)$ off-shell:
\begin{eqnarray}\label{Mixedcond1}
\Pi_{nonper1}^{D(B)}&=&-\langle\overline{s}s\rangle\Big[\frac{m_{0}^2}{24}
\Big(\frac{m_{c(b)}^2-q^2}{r^{2}r'}+\frac{2}
{rr'}+\frac{1}{r^2}+\frac{1}{r'^2}+
\frac{m_{c(b)}^2-q^2}{rr^{'2}}\Big)
\nonumber\\
&+&\frac{q^2+2m_{u}m_{c(b)}-m_{c(b)}^2}{2rr'}-\frac{1}{2r}-\frac{1}{r'}\Big],\nonumber\\
\end{eqnarray}
\begin{eqnarray}\label{Mixedcond12}
\Pi_{nonper2}^{D(B)}&=&-\langle\overline{s}s\rangle \Big[
\frac{m_{c(b)}}{rr'}
-\frac{m_{0}^2~m_{c(b)}}{24~rr'^2}\Big],\nonumber\\
\end{eqnarray}
\item $K_1$ off-shell:
\begin{eqnarray}\label{CorrelationFuncNonpertByy}
\Pi_{nonper1}^{K_1}&=&0,
\end{eqnarray}
\begin{eqnarray}\label{CorrelationFuncNonpertByy2}
\Pi_{nonper2}^{K_1}&=&0,
\end{eqnarray}
\end{itemize}
\end{itemize}
where $r={p}^2-m_{c(b)}^2$ and $r'={p^{\prime}}^2-m_u^2$. The $\Pi_{nonper1}$ and  $\Pi_{nonper2}$ correspond to also the currents $\eta_\nu^{K_1}$ and $\eta_{\nu\nu'}^{K_1}$, respectively. In this step, we equate the physical side in the case of $K_0^*$ and the coefficients of the selected structures in physical sides of $K$ 
and $K_1$ to the corresponding QCD sides and obtain QCD sum rules for the considered strong coupling constants. To suppress the contributions of the higher states and continuum, we also apply the double Borel
 transformation with respect to the variables $p^2(p^2\longrightarrow M^2)$ and
$p^{\prime^2}(p^{\prime^2}\longrightarrow M^{\prime^2})$. Finally, we
get the following sum rules for the considered coupling constants:
\begin{itemize}
 \item  $D_{s}(B_{s})-D(B)-K_0^*$ vertex:
\begin{itemize}
\item $D(B)$ off-shell:
\begin{eqnarray}\label{CoupCons-gDsDKs-DBoff}
&&g^{D(B)}_{D_{s}DK^{\ast}_0(B_{s}BK^{\ast}_0)}(q^2)=\frac{2(q^2-m_{D(B)}^2)
(m_{c(b)}+m_u)(m_{c(b)}+m_s)}{m_{D_s(B_s)}^2m_{D(B)}^2m_{K^{\ast}_0}
f_{D_s(B_s)}f_{D(B)}
f_{K^{\ast}_0}(m_{D_s(B_s)}^2+m_{K^{\ast}_0}^2-q^2)}e^{\frac{m_{D_s(B_s)}^2}{M^2}}
\nonumber \\
&\times&e^{\frac{m_{K^{\ast}_0}^2}{{M^{\prime}}^2}}\Big[-\frac{1}{4\pi^2}\int^{s_0}_{(m_{c(b)}+m_s)^2}
ds\int^{s^{\prime}_0}_{(m_s+m_u)^2} ds^{\prime}
\rho^{D(B)}(s,s^{\prime},q^2)\theta
[1-{(f^{D(B)}(s,s^{\prime}))}^2]e^{\frac{-s}{M^2}}e^{\frac{-s^{\prime}}
{{M^{\prime}}^2}}\nonumber \\&+&\widehat{B}\Pi_{nonper}^{D(B)}\Big],
\end{eqnarray}
\item $K_0^*$ off-shell:
\begin{eqnarray}\label{CoupCons-gDsDKs-KBoff}
&&g^{K^{\ast}_0}_{D_{s}DK^{\ast}_0(B_{s}BK^{\ast}_0)}(q^2)=\frac{-2
(q^2-m_{K^{\ast}_0}^2)
(m_{c(b)}+m_u)(m_{c(b)}+m_s)}{m_{D_s(B_s)}^2m_{D(B)}^2m_{K^{\ast}_0}
f_{D_s(B_s)}f_{D(B)}
f_{K^{\ast}_0}(m_{D_s(B_s)}^2+m_{D(B)}^2-q^2)}e^{\frac{m_{D_s(B_s)}^2}{M^2}}
\nonumber \\
&\times&e^{\frac{m_{D_s(B_s)}^2}{{M^{\prime}}^2}} \Big[-\frac{1}{4\pi^2}\int^{s_0}_{(m_{c(b)}+m_s)^2}
ds\int^{s^{\prime}_0}_{(m_{c(b)}+m_u)^2} ds^{\prime}
\rho^{K^{\ast}_0}(s,s^{\prime},q^2) \theta
[1-{(f^{K^{\ast}_0}(s,s^{\prime}))}^2]e^{\frac{-s}{M^2}}e^{\frac{-s^{\prime}}
{{M^{\prime}}^2}}\Big],\nonumber \\
\end{eqnarray}
\end{itemize}
\item  $D^*_{s}(B^*_{s})-D(B)-K$ vertex:
\begin{itemize}
\item $D(B)$ off-shell:
\begin{eqnarray}\label{CoupCons-gDsDKs-Doffshel}
&&g^{D(B)}_{D^{\ast}_{s}DK(B^{\ast}_{s}BK)}(q^2)=\frac{(q^2-m_{D(B)}^2)(m_{c(b)}+m_u)(m_s+m_u)}
{f_{D_s^{\ast}(B_s^{\ast})} f_{D(B)} f_{K}m_{D_s^{\ast}(B_s^{\ast})}m_{K}^2 m_{D(B)}^2(1+
\frac{m_K^2-q^2}{m_{D_s^{\ast}(B_s^{\ast})}^2})}
e^{\frac{m_{D_s^{\ast}(B_s^{\ast})}^2}{M^2}}e^{\frac{m_{K}^2}{{M^{\prime}}^2}}
\nonumber \\
&\times&\Bigg[-\frac{1}{4~\pi^2}\int^{s_0}_{(m_{c(b)}+m_s)^2}
ds\int^{s^{\prime}_0}_{(m_s+m_u)^2} ds^{\prime}
\rho^{D(B)}(s,s^{\prime},q^2) \theta
[1-{(f^{D(B)}(s,s^{\prime}))}^2]e^{\frac{-s}{M^2}}e^{\frac{-s^{\prime}}
{{M^{\prime}}^2}}\nonumber \\&+&\widehat{B}\Pi_{nonper}^{D(B)}\Bigg]
\end{eqnarray}
\item $K$ off-shell:
\end{itemize}
\begin{eqnarray}\label{CoupCons-gDsDKs-Ksoffshel}
&&g^{K}_{D^{\ast}_{s}DK(B^{\ast}_{s}BK)}(q^2)=\frac{(q^2-m_K^2)(m_{c(b)}+m_u)(m_s+m_u)}
{f_{D_s^{\ast}(B_s^{\ast})} f_{D(B)} f_{K}m_{D_s^{\ast}(B_s^{\ast})}m_{K}^2 m_{D(B)}^2(1+
\frac{m_{D(B)}^2-q^2}{m_{D_s^{\ast}(B_s^{\ast})}^2})}
e^{\frac{m_{D_s^{\ast}(B_s^{\ast})}^2}{M^2}}e^{\frac{m_{D(B)}^2}{{M^{\prime}}^2}}
\nonumber \\
&\times&\left[-\frac{1}{4~\pi^2}\int^{s_0}_{(m_{c(b)}+m_s)^2}
ds\int^{s^{\prime}_0}_{(m_{c(b)}+m_u)^2} ds^{\prime}
\rho^{K}(s,s^{\prime},q^2) \right.
 \left. \theta
[1-{(f^{K}(s,s^{\prime}))}^2]e^{\frac{-s}{M^2}}e^{\frac{-s^{\prime}}
{{M^{\prime}}^2}}\right],\nonumber\\
\end{eqnarray}
\item  $D^*_{s}(B^*_{s})-D(B)-K_1$ vertex:
\begin{itemize}
\item $D(B)$ off-shell:
\begin{eqnarray}\label{CoupCons-gDsDK1A-Doffshel}
&&\Bigg(g^{D(B)}_{D^{\ast}_{s}D
K_1^{a}(B^{\ast}_{s}B K_1^{a})}(q^2)m_{K^{a}_1}
f_{K^{a}_1}e^{\frac{-m_{K_1^{a}}^2}{{M^{\prime}}^2}}+g^{D(B)}_{D^{\ast}_{s}D
K_1^{b}(B^{\ast}_{s}B K_1^{b})}(q^2)m_{K^b_1}
f_{K^{b\bot}_1}a_0^{\parallel,K^{b}_1}e^{\frac{-m_{K_1^{b}}^2}{{M^{\prime}}^2}}\Bigg)\nonumber\\&=&\frac{(q^2-m_{D(B)}^2)}{f_{D_s^{\ast}(B_s^{\ast})}
f_{D(B)}
\frac{m_{D(B)}^2}{m_{c(b)}+m_u}m_{D_s^{\ast}(B_s^{\ast})}^3} e^{\frac{m_{D^{\ast}_s(B_s^{\ast})}^2}{M^2}}
\nonumber \\
&\times&
\left[-\frac{1}{4~\pi^2}\int^{s_0}_{(m_{c(b)}+m_s)^2}
ds\int^{s^{\prime}_0}_{(m_s+m_u)^2} ds^{\prime}
\rho^{D(B)}_1(s,s^{\prime},q^2) \theta
[1-{(f^{D(B)}(s,s^{\prime}))}^2]\right.\nonumber \\
&\times& \left.e^{\frac{-s}{M^2}}e^{\frac{-s^{\prime}}
{{M^{\prime}}^2}}+\widehat{B}\Pi_{nonper1}^{D(B)}\right],
\end{eqnarray}
\begin{eqnarray}\label{CoupCons-gDsDK1A-Doffshel2}
&&\Bigg(g^{D(B)}_{D^{\ast}_{s}D
K_1^{a}(B^{\ast}_{s}B K_1^{a})}(q^2)f_{K_{1}^{a}}~a_{0}^{\perp,K_{1}^{a}}e^{\frac{-m_{K_1^{a}}^2}{{M^{\prime}}^2}}+g^{D(B)}_{D^{\ast}_{s}D
K_1^{b}(B^{\ast}_{s}B K_1^{b})}(q^2)
f_{K^{b\bot}_1}e^{\frac{-m_{K_1^{b}}^2}{{M^{\prime}}^2}}\Bigg)\nonumber\\&=&\frac{(q^2-m_{D(B)}^2)}{f_{D_s^{\ast}(B_s^{\ast})}
f_{D(B)}
\frac{m_{D(B)}^2}{m_{c(b)}+m_u}m_{D_s^{\ast}(B_s^{\ast})}^3} e^{\frac{m_{D^{\ast}_s(B_s^{\ast})}^2}{M^2}}
\nonumber \\
&\times&
\left[-\frac{1}{4~\pi^2}\int^{s_0}_{(m_{c(b)}+m_s)^2}
ds\int^{s^{\prime}_0}_{(m_s+m_u)^2} ds^{\prime}
\rho^{D(B)}_2(s,s^{\prime},q^2) \theta
[1-{(f^{D(B)}(s,s^{\prime}))}^2]\right.\nonumber \\
&\times& \left.e^{\frac{-s}{M^2}}e^{\frac{-s^{\prime}}
{{M^{\prime}}^2}}+\widehat{B}\Pi_{nonper2}^{D(B)}\right],
\end{eqnarray}
\item $K_1$ off-shell:
\begin{eqnarray}\label{CoupCons-gDsDK1A-Doffshel}
&&\Bigg(g^{K^{a}_1}_{D^{\ast}_{s}D
K_1^{a}(B^{\ast}_{s}B K_1^{a})}(q^2)\frac{m_{K^{a}_1}
f_{K^{a}_1}}{(q^2-m_{K^{a}_1}^2)}+g^{K^{b}_1}_{D^{\ast}_{s}D
K_1^{b}(B^{\ast}_{s}B K_1^{b})}(q^2)\frac{m_{K^b_1}
f_{K^{b\bot}_1}a_0^{\parallel,K^{b}_1}}{(q^2-m_{K^{b}_1}^2)}\Bigg)\nonumber\\&=&\frac{1}{f_{D_s^{\ast}(B_s^{\ast})}
f_{D(B)}
\frac{m_{D(B)}^2}{m_{c(b)}+m_u}m_{D_s^{\ast}(B_s^{\ast})}^3} e^{\frac{m_{D^{\ast}_s(B_s^{\ast})}^2}{M^2}}e^{\frac{m_{D(B)}^2}{{M^{\prime}}^2}}
\nonumber \\
&\times&
\left[-\frac{1}{4~\pi^2}\int^{s_0}_{(m_{c(b)}+m_s)^2}
ds\int^{s^{\prime}_0}_{(m_{c(b)}+m_u)^2} ds^{\prime}
\rho^{K_1}_1(s,s^{\prime},q^2) \theta
[1-{(f^{K_1}(s,s^{\prime}))}^2]\right.\nonumber \\
&\times& \left.e^{\frac{-s}{M^2}}e^{\frac{-s^{\prime}}
{{M^{\prime}}^2}}+\widehat{B}\Pi_{nonper1}^{K_1}\right],
\end{eqnarray}
\begin{eqnarray}\label{CoupCons-gDsDK1A-Doffshel2}
&&\Bigg(g^{K_{1}^{a}}_{D^{\ast}_{s}D
K_1^{a}(B^{\ast}_{s}B K_1^{a})}(q^2)\frac{f_{K_{1}^{a}}~a_{0}^{\perp,K_{1}^{a}}}{(q^2-m_{K_1^{a}}^2)}+g^{K_{1}^{b}}_{D^{\ast}_{s}D
K_1^{b}(B^{\ast}_{s}B K_1^{b})}(q^2)
\frac{f_{K^{b\bot}_1}}{(q^2-m_{K_1^{b}}^2)}\Bigg)\nonumber\\&=&\frac{1}{f_{D_s^{\ast}(B_s^{\ast})}
f_{D(B)}
\frac{m_{D(B)}^2}{m_{c(b)}+m_u}m_{D_s^{\ast}(B_s^{\ast})}^3} e^{\frac{m_{D^{\ast}_s(B_s^{\ast})}^2}{M^2}}e^{\frac{m_{D(B)}^2}{{M^{\prime}}^2}}
\nonumber \\
&\times&
\left[-\frac{1}{4~\pi^2}\int^{s_0}_{(m_{c(b)}+m_s)^2}
ds\int^{s^{\prime}_0}_{(m_{c(b)}+m_u)^2} ds^{\prime}
\rho^{K_1}_2(s,s^{\prime},q^2) \theta
[1-{(f^{K_1}(s,s^{\prime}))}^2]\right.\nonumber \\
&\times& \left.e^{\frac{-s}{M^2}}e^{\frac{-s^{\prime}}
{{M^{\prime}}^2}}+\widehat{B}\Pi_{nonper2}^{K_1}\right],
\end{eqnarray}

\end{itemize}
\end{itemize}
where the $\widehat{B}\Pi_{nonper}$ represents the double Borel
transformation of the non-perturbative part in each case, $s_0$ and $s_0^{\prime}$ are the continuum thresholds and the functions $f^i(s,s^{\prime})$ inside the step functions
are determined requiring that the arguments of the three $\delta$
functions coming from the Cutkosky rule vanish simultaneously. As a result, we find:
\begin{itemize}
\item $D(B)$ off-shell:
\begin{eqnarray}\label{fsspD0offshell}
f^{D(B)}(s,s^{\prime})=\frac{2~s~(m_s^2-m_u^2+s^{\prime})+(m_{c(b)}^2-m_s^2-s)
(-q^2+s+s^{\prime})}{\lambda^{1/2}(m_{c(b)}^2,m_s^2,s)
\lambda^{1/2}(s,s^{\prime},q^2)},
\end{eqnarray}
\item $K_0^*(K)(K_1)$ off-shell:
\begin{eqnarray}\label{fsspKs1offshell}
f_1^{K_0^*(K)(K_1)}(s,s^{\prime})=\frac{2~s~(-m_{c(b)}^2+m_u^2-s^{\prime})+(m_{c(b)}^2-m_s^2+s)
(-q^2+s+s^{\prime})}{\lambda^{1/2}(m_{c(b)}^2,m_s^2,s)
\lambda^{1/2}(s,s^{\prime},q^2)}.\nonumber\\
\end{eqnarray}
\end{itemize}
Here, we should stress that the physical regions are imposed
by the limits on the integrals and step functions  in the integrands
in the   sum rules expressions. In order to subtract the contributions of the
higher states and continuum, the quark-hadron duality assumption in the following form is
used:
\begin{eqnarray}\label{ope}
\rho^{higher states}(s, s')=\rho^{OPE}(s, s')\theta(s-s_0)
\theta(s'-s'_0).
\end{eqnarray}
The double Borel transformation used in calculations is also
defined in the following way:
\begin{equation}\label{16au}
\hat{B}\frac{1}{(p^2-m^2_1)^m}\frac{1}{(p'^2-m^2_2)^n}\rightarrow(-1)^{m+n}\frac{1}{\Gamma(m)}\frac{1}{\Gamma
(n)}e^{-m_{1}^2/M^{2}}e^{-m_{2}^2/M^{'2}}\frac{1}{(M^{2})^{m-1}(M^{'2})^{n-1}}.
\end{equation}

At the end of this section, we would like to mention that using Eqs. (\ref{melo}) and (\ref{transitionamp}), the couplings to $K_1(1270)$ and  $K_1(1400)$ are obtained in terms of the couplings to the $K_1^{a(b)}$ as:
\begin{eqnarray}\label{coupling constant sumrule}
g_{D^{\ast}_{s}DK_{1}(1270)(B^{\ast}_{s}BK_{1}(1270))}&=&g_{D^{\ast}_{s}DK_1^a(B^{\ast}_{s}BK_1^a)}\sin
\theta+g_{D^{\ast}_{s}DK_1^b(B^{\ast}_{s}BK_1^b)}\cos \theta\nonumber\\
g_{D^{\ast}_{s}DK_{1}(1400)(B^{\ast}_{s}BK_{1}(1400))}&=&g_{D^{\ast}_{s}DK_1^a(B^{\ast}_{s}BK_1^a)}\cos
\theta-g_{D^{\ast}_{s}DK_1^b(B^{\ast}_{s}BK_1^b)}\sin \theta
\end{eqnarray}

\section{Numerical analysis}

In the present section, we numerically analyze the expressions of QCD sum rules obtained for the considered strong coupling constants. Some input parameters used in the calculations are: $m_{K}=
 (493.677\pm0.016)~MeV$, $m_{K^{\ast}_0}(800)=(672\pm40)~MeV$,
$m_{K^{\ast}_0}(1430)=(1425\pm50)~MeV$, $m_{K_1}(1270)=(1272\pm7)~MeV$, $m_{K_1}(1400)=(1403\pm7)~MeV$,  $m_{D}=(1.8648\pm0.00014)~GeV$, $m_{B}=(5.2792\pm
 0.0003)~GeV$,
$m_{D_s}=(1.96847\pm0.00033)~MeV$, $m_{B_s}=(5.3663\pm0.0006)~MeV$, 
 $m_{D_{s}^{\ast}}=(2.1123\pm0.0005)~GeV$, $m_{B_s^{\ast}}=(5.4154\pm0.0014)~GeV$ \cite{K.Nakamura},
$m_c=1.3~GeV$, $m_{b}=4.7~GeV$, $m_s=0.14~GeV$\cite{B.L.Ioffe}, $f_{K}=160~MeV$\cite{S.Eidelman},
$f_{K^{\ast}_0(800)}(1~GeV)=(340\pm20)~MeV$,
$f_{K^{\ast}_0(1430)}(1~GeV)=(445\pm50)~MeV$ \cite{H-Y.Cheng},
$f_{D_s^{\ast}}=(272\pm16^{0}_{-20})~MeV$,
$f_{B_s^{\ast}}=(229\pm20^{31}_{-16})~MeV$ \cite{D.Becirevic},
$f_{B}=190\pm13~MeV$ \cite{E.Gamiz}, $f_D=(202\pm41\pm17)~MeV$ \cite{I.Danko},  $f_{D_s}=(286\pm44\pm41)~MeV$
\cite{G.Abbiendi}, $f_{B_s}=196~MeV$ \cite{M.A.Ivanov}, ${\langle}\overline{s}s{\rangle}=0.8{\langle}\overline{u}u{\rangle}=-0.8(0.24\pm0.01)^3~GeV^3$
\cite{B.L.Ioffe}, $m_0^2=(0.8\pm0.2)~GeV^2$
  \cite{VMBelyaev}, $m_{K_1^a}=1.31~GeV$, 
$m_{K_1^b}=1.34~GeV$, 
$f_{K_1^a}(1
~GeV)=0.25~GeV$,
$f_{K_1^{b\perp}}(1~GeV)=0.19~GeV$,
$a_0^{\|,K_1^b}(1
~GeV)=-0.19\pm0.07 $ and $a_{0}^{\perp,K_{1}^{A}}(1
~GeV)=0.27^{+0.03} _{-0.17}$ \cite{lee,hatanaka,busey}.

 The sum rule for strong coupling constants also contain  four auxiliary
parameters, namely the continuum thresholds $s_0$ and
$s_0^{\prime}$ related to the initial and final channels, respectively as well as Borel mass parameters $M^2$ and
${M^{\prime}}^2$. These 
quantities are mathematical objects, so according to the general criteria and standard procedure in QCD sum rules,   our physical results should  be insensitive to them. Therefore,
we shall look for working regions of these quantities  at which the dependence of coupling
constants on these auxiliary parameters are weak. The working
regions for the Borel mass parameters $M^2$ and $M^{\prime^2}$ are
determined demanding that both the contributions of the higher
states and continuum are sufficiently suppressed and the
contributions coming from the higher dimensional operators are small. Our calculations lead to the following working regions common for all cases:
\begin{itemize}
 \item $D_s^{(\ast)}D K_0^*(K)(K_1)$ vertex:
\begin{itemize} \item  $D$ off-shell: $8 GeV^2\leq M^2 \leq 25
GeV^2$ and $5 GeV^2 \leq M^{\prime^2} \leq 15 GeV^2$,  
\item $K_0^*(K)(K_1)$
off-shell: $6 GeV^2 \leq M^2 \leq 15 GeV^2$ and $4 GeV^2 \leq
M^{\prime^2} \leq 12GeV^2$,
\end{itemize}
\item $B_s^{(\ast)}B K_0^*(K)(K_1)$ vertex:
\begin{itemize}
\item  $B$ off-shell: $14 GeV^2\leq M^2 \leq 30 GeV^2$ and
$5 GeV^2 \leq M^{\prime^2} \leq 20 GeV^2$, 

\item  $K_0^*(K)K_1$ off-shell: $6
GeV^2\leq M^2 \leq20 GeV^2$ and $5 GeV^2 \leq M^{\prime^2} \leq 15
GeV^2$.
\end{itemize}
\end{itemize}
 The continuum
thresholds, $s_0$ and $s^{\prime}_0$ are not completely arbitrary
but they are correlated to the energy of the first excited states
with the same quantum numbers as the considered interpolating currents. Our numerical calculations show that in
the  regions 
$(m_{i}+0.3)^2\leq s_0
\leq(m_{i}+0.7)^2$ and $(m_{f}+0.3)^2\leq
s_0^{\prime}\leq (m_{f}+0.7)^2$, respectively  for the continuum thresholds $s$ and $s'$, our results have weak dependence on these parameters. Here, $m_i$ is the mass of initial particle and the $m_f$
 stands for the 
mass of the final on-shell state. For instance consider the $g^{D}_{D_{s}DK^{\ast}_0}$ coupling constant at which the $D$ meson is off-shell. This coupling constant describe the strong 
transition $D_{s}\rightarrow DK^{\ast}_0$, and for this case
$m_i=m_{D_s}$ and $m_f=m_{K^{\ast}_0}$.

\begin{figure}[h!]
\includegraphics[totalheight=6cm,width=7cm]{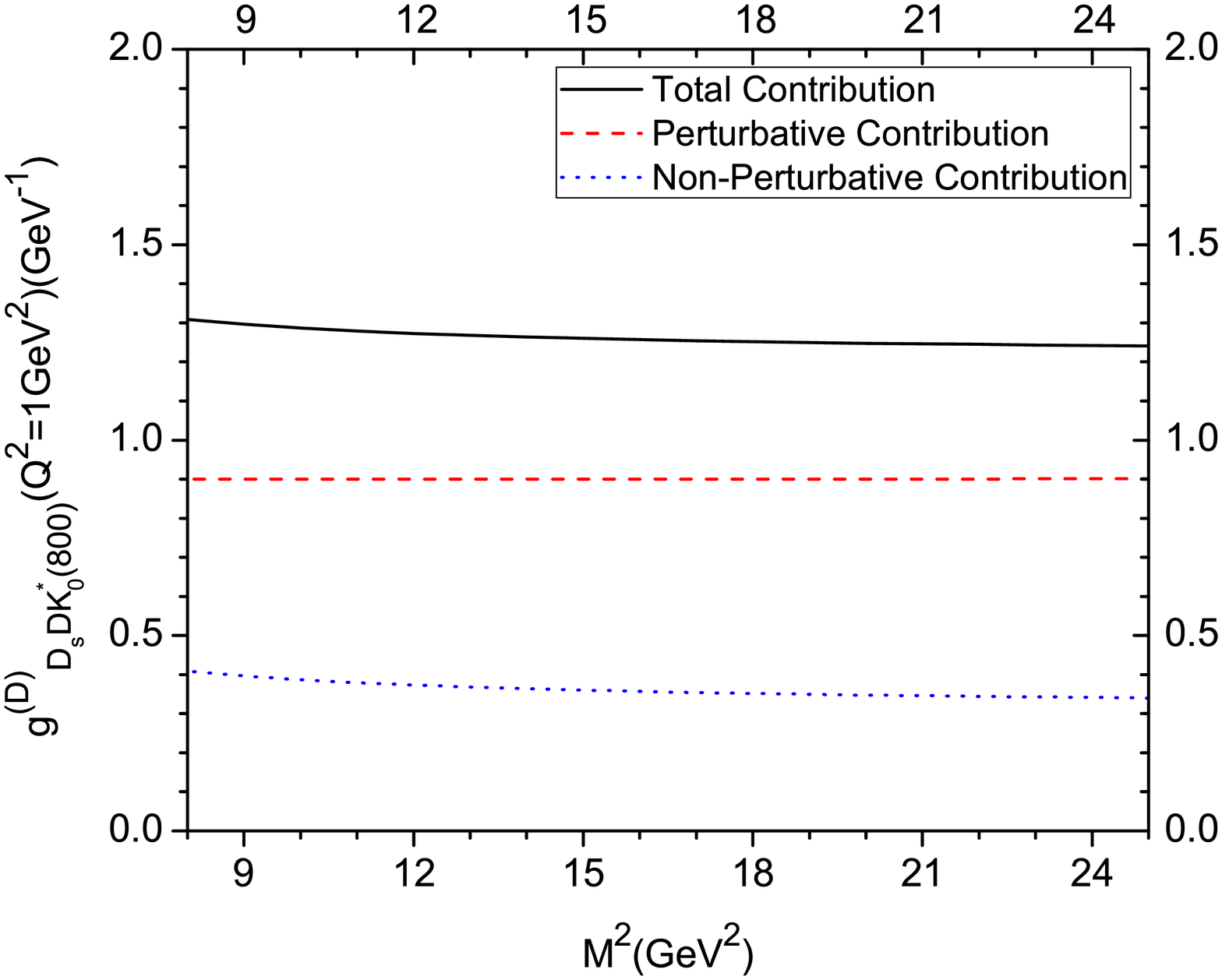}
\includegraphics[totalheight=6cm,width=7cm]{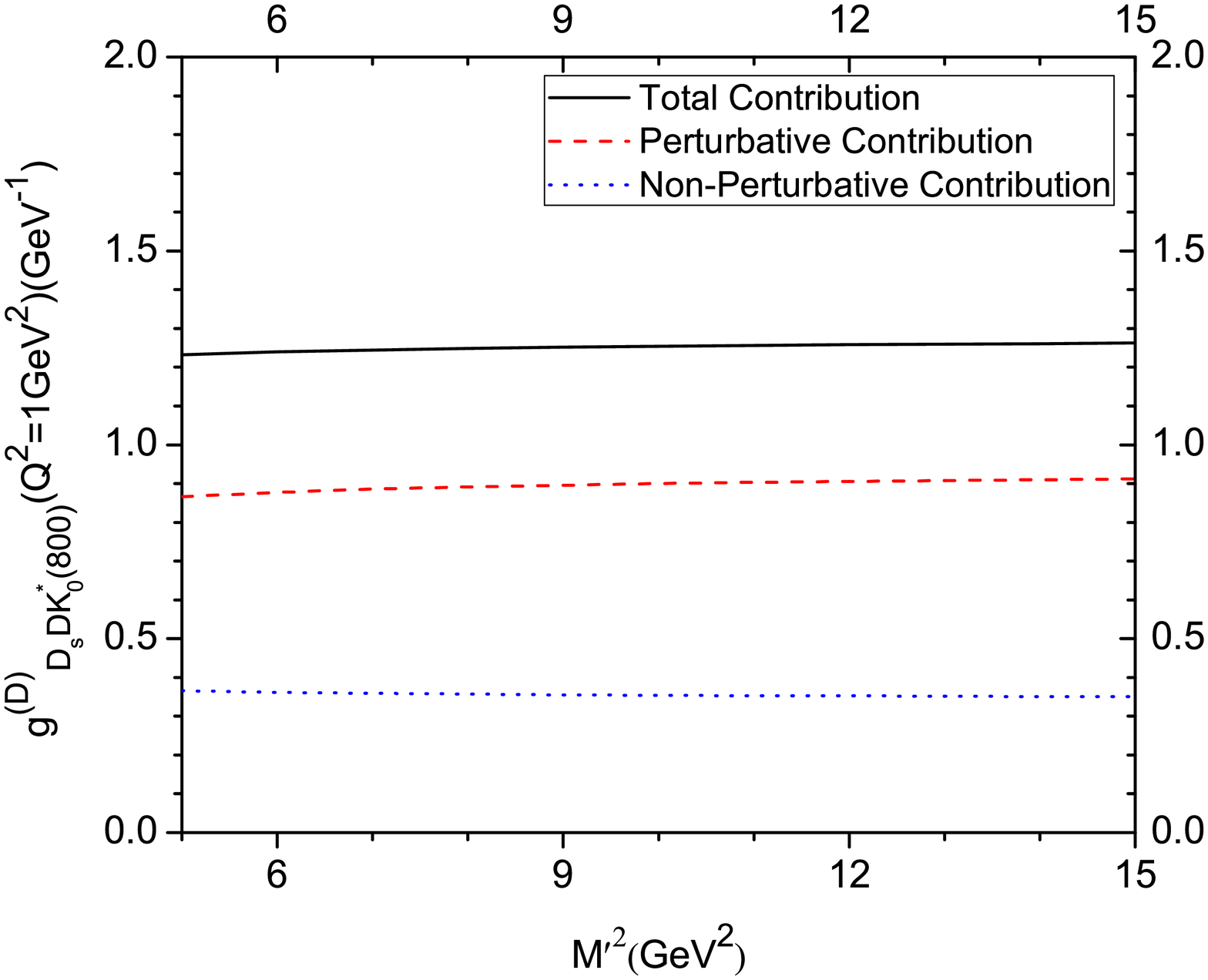}
\caption{\textbf{Left:} $g^{(D)}_{D_sDK^*_0(800)}(Q^2=1~GeV^2)$ as a
function of the Borel mass $M^2$ with ${M^{\prime}}^2=10~GeV^2$.
\textbf{Right:} $g^{(D)}_{D_sDK^*_0(800)}(Q^2=1~GeV^2)$ as a
function of the Borel mass ${M^{\prime}}^2$ with $M^2=17~GeV^2$. The
continuum thresholds $s_0=6.09~GeV^2$ and $s_0^{\prime}=1.37~GeV^2$
have been used. } \label{gDsDK800Doff}
\end{figure}

As an example, we present the  dependence of strong coupling constant $g^{(D)}_{D_sDK^*_0(800)}$  on Borel mass parameters
at $Q^2=1~GeV^2$, where $Q^2=-q^2$ in 
Fig. (\ref{gDsDK800Doff}). This figure demonstrates a good stability of the results with respect
to the variations of Borel mass parameters in their working regions.

Now, we proceed to find the $Q^2$ behavior of the considered strong coupling constants using the working regions for auxiliary parameters.  First, we consider the scalar kaon case for both $K^*_0(800)$ and
$K^*_0(1430)$. The
strong coupling constant  in this case obeys from the following Boltzmann function:
\begin{eqnarray}\label{gFitQsq}
g(Q^2)=A_1+\frac{A_2}{1+\exp[\frac{Q^2-x_0}{\Delta x}]}~[GeV^{-1}].
\end{eqnarray}
The values of the parameters $A_1$, $A_2$, $x_0$ and $\Delta x$ for  the considered 
coupling constant form factors are given in Table \ref{fitparametrization0}.
\begin{table}[h]
\renewcommand{\arraystretch}{1.5}
\addtolength{\arraycolsep}{3pt}
$$
\begin{array}{|c|c|c|c|c|}
\hline \hline
 \mbox{ } &
 A_1 &  A_2&  x_0
 &  \Delta x
 \\
\hline\hline
  \mbox{$g^{(D)}_{D_sDK^*_0(800)}(Q^2)$}
      & 3.468 & -2.741 & 8.067&4.995
      \\
      \hline
  \mbox{$g^{(K^*_0(800))}_{D_sDK^*_0(800)}(Q^2)$}
      & -0.024 & 0.772 & 5.723&1.257
      \\
      \hline
  \mbox{$g^{(D)}_{D_sDK^*_0(1430)}(Q^2)$}
      & 4.712 & -3.818 & 24.863&10.985
      \\
       \hline
  \mbox{$g^{(K^*_0(1430))}_{D_sDK^*_0(1430)}(Q^2)$}
      & -0.022 & 0.772 & 4.729&1.637
      \\
      \hline
  \mbox{$g^{(B)}_{B_sBK^*_0(800)}(Q^2)$}
      & 4.151 & -1.932 & 13.842&12.149
      \\
       \hline
  \mbox{$g^{(K^*_0(800))}_{B_sBK^*_0(800)}(Q^2)$}
      & -0.017 & 0.547 & 5.431&1.121
      \\
       \hline
  \mbox{$g^{(B)}_{B_sBK^*_0(1430)}(Q^2)$}
      & 2.055 & -0.207 & 11.239&5.084
      \\
      \hline
  \mbox{$g^{(K^*_0(1430))}_{B_sBK^*_0(1430)}(Q^2)$}
      & -0.004 & 0.255 & 4.819&1.146
      \\
    \hline \hline
\end{array}
$$
\caption{Parameters appearing in the fit function of the
coupling constants for $D_sDK^*_0(800)$, $D_sDK^*_0(1430)$,
$B_sBK^*_0(800)$ and $B_sBK^*_0(1430)$ vertices. $A_1$ and $A_2$ are in $GeV^{-1}$ units, while $x_0$ and $\Delta x$ are in the units of $GeV^2$.}
\label{fitparametrization0}
\renewcommand{\arraystretch}{1}
\addtolength{\arraycolsep}{-1.0pt}
\end{table}
The coupling constants are defined as the values of the form factors
at $Q^2=-m_{meson}^2$, where $m_{meson
}$ is the mass of  off-shell meson. The results of the coupling
constants obtained using $Q^2=-m_{meson}^2$ are given in Tables
\ref{gDsDK1430} and \ref{gBsBK1430}. The final result for each coupling constant is obtained taking the average of the coupling constants obtained from two different off-shell cases. 
The errors in the numerical results are due to the uncertainties in
determination of the working regions for the auxiliary parameters as
well as the errors in the input parameters.
\begin{table}[h]
\renewcommand{\arraystretch}{1.5}
\addtolength{\arraycolsep}{3pt}
$$
\begin{array}{|c|c|c|c|}
\hline \hline
 \mbox{ } &
 Q^2=-m_D^2 &  Q^2=-m_{K^*_0(800)}^2&  \mbox{Average}

 \\
\hline
  \mbox{$g_{D_sDK^*_0(800)}$}
      & 0.97\pm0.02 & 0.74\pm0.05 & 0.85\pm0.08
      \\
    \hline \hline
 \mbox{ } &
 Q^2=-m_D^2 &  Q^2=-m_{K^*_0(1430)}^2&  \mbox{Average}

 \\
\hline
  \mbox{$g_{D_sDK^*_0(1430)}$}
      & 1.16\pm0.12 & 0.49\pm0.07 & 0.83\pm0.09
      \\
    \hline \hline
\end{array}
$$
\caption{Value of the $g_{D_sDK^*_0(800,1430)}$ coupling constant in
$GeV^{-1}$ unit.} \label{gDsDK1430}
\renewcommand{\arraystretch}{1}
\addtolength{\arraycolsep}{-1.0pt}
\end{table}
\begin{table}[h]
\renewcommand{\arraystretch}{1.5}
\addtolength{\arraycolsep}{3pt}
$$
\begin{array}{|c|c|c|c|}
\hline \hline
 \mbox{ } &
 Q^2=-m_B^2 &  Q^2=-m_{K^*_0(800)}^2&  \mbox{Average}

 \\
\hline
  \mbox{$g_{B_sBK^*_0(800)}$}
      & 2.28\pm0.18 & 0.53\pm0.09 & 1.41\pm0.21
      \\
    \hline \hline
 \mbox{ } &
 Q^2=-m_B^2 &  Q^2=-m_{K^*_0(1430)}^2& \mbox{Average}

 \\
\hline
  \mbox{$g_{B_sBK^*_0(1430)}$}
      & 1.85\pm0.53 & 0.25\pm0.04 & 1.05\pm0.32
      \\
    \hline \hline
\end{array}
$$
\caption{Value of the $g_{B_sBK^*_0(800,1430)}$ coupling constant in
$GeV^{-1}$ unit.} \label{gBsBK1430}
\renewcommand{\arraystretch}{1}
\addtolength{\arraycolsep}{-1.0pt}
\end{table}

 In the case
of pseudoscalar kaon and $D$ off-shell, the strong coupling constant is well described  by the following monopolar fit
parametrization:
\begin{eqnarray}\label{DsDKFitDoffShell}
g^{(D)}_{D^{\ast}_sDK}(Q^2)=\frac{8.76~(GeV^2)}{Q^2+7.12~(GeV^2)},
\end{eqnarray}
 The value of coupling constant obtained  at $Q^2 = -m^2_{meson}$ is presented in  Table (\ref{kazem1}).
\begin{table}[h]
\renewcommand{\arraystretch}{1.5}
\addtolength{\arraycolsep}{3pt}
$$
\begin{array}{|c|c|c|c|}
\hline \hline
 \mbox{ } &
 Q^2=-m_D^2 &  Q^2=-m_{K}^2&  \mbox{Average}

 \\
\hline\hline
  \mbox{$g_{D^*_sDK}$ (Present work)}
      & 2.79\pm0.24 & 2.99\pm0.26 & 2.89\pm0.25
      \\
\hline
  \mbox{$g_{D^*_sDK}$ (\cite{M.E.Bracco})}
      & 2.72 & 2.87 & 2.84\pm0.31
      \\
    \hline \hline
\end{array}
$$
\caption{Value of the $g_{D^*_sDK}$ coupling constant. }
\label{kazem1}
\renewcommand{\arraystretch}{1}
\addtolength{\arraycolsep}{-1.0pt}
\end{table}

The result for   strong coupling constant of pseudoscalar case and an
off-shell $K$ meson can be well fitted by the exponential
parametrization
\begin{eqnarray}\label{DsDKFitKoffShell}
g^{(K)}_{D^{\ast}_sDK}(Q^2)=3.55 ~e^{-\frac{Q^2}{7.25~(GeV^2)}}-0.88,
\end{eqnarray}
where using $Q^2=-m_{K}^2$, we obtain the result as also presented in the Table (\ref{kazem1}). We also depict the final result for this case taking the average
of two above obtained values. This Table also shows the predictions of  \cite{M.E.Bracco} on  $g_{D^{\ast}_sDK}$ as the only existing previously calculated coupling constant among the 
considered vertices. Comparing our results with that of \cite{M.E.Bracco}, we see a good consistency between two predictions.

Similarly, for $B^{\ast}_sBK$ vertex, our result for the pseudoscalar kaon and $B$ off-shell is
better extrapolated by the exponential fit parametrization
\begin{eqnarray}\label{BsBKFitBoffShell}
g^{(B)}_{B^{\ast}_sBK}(Q^2)=0.66~ e^{-\frac{Q^2}{23.34~(GeV^2)}}+0.23,
\end{eqnarray}
and in the  $K$ off-shell case by  the parametrization
\begin{eqnarray}\label{BsBKFitKoffShell}
g^{(K)}_{B^{\ast}_sBK}(Q^2)=4.39 ~e^{-\frac{Q^2}{4.02~(GeV^2)}}-1.03.
\end{eqnarray}
Using the same procedure as above, we find the values depicted in the Table (\ref{kazem2}).
\begin{table}[h]
\renewcommand{\arraystretch}{1.5}
\addtolength{\arraycolsep}{3pt}
$$
\begin{array}{|c|c|c|c|}
\hline \hline
 \mbox{ } &
 Q^2=-m_B^2 &  Q^2=-m_{K}^2&  \mbox{Average}

 \\
\hline\hline
  \mbox{$g_{B^*_sBK}$}
      & 2.40\pm0.22 & 3.62\pm0.34 & 3.01\pm0.28
      \\
    \hline \hline
\end{array}
$$
\caption{Value of the $g_{B^*_sBK}$ coupling constant. }
\label{kazem2}
\renewcommand{\arraystretch}{1}
\addtolength{\arraycolsep}{-1.0pt}
\end{table}

In the case of axial vector kaon, the strong coupling constant  obey also the same 
 Boltzmann function as the scalar case.
The values of the parameters $A_1$, $A_2$, $x_0$ and $\Delta x$ for
coupling constants in this case are given in Table \ref{fitparametrization}.
\begin{table}[h]
\renewcommand{\arraystretch}{1.5}
\addtolength{\arraycolsep}{3pt}
$$
\begin{array}{|c|c|c|c|c|}
\hline \hline
 \mbox{ } &
 A_1 &  A_2&  x_0
 &  \Delta x
 \\
\hline\hline
  \mbox{$g^{(D)}_{D_s^*DK_{1}(1270)}(Q^2)$}
      & 5.062 & -2.337 & 1.182&1.531
      \\
      \hline
  \mbox{$g^{(D)}_{D_s^*DK_{1}(1400)}(Q^2)$}
      & 73.848 & -87.162 &118.101&74.590
      \\
      \hline
  \mbox{$g^{(K_{1}(1270))}_{D_s^*DK_{1}(1270)}(Q^2)$}
      & 0.137 & -1.507 & 6.951&1.845
      \\
       \hline
  \mbox{$g^{(K_{1}(1400))}_{D_s^*DK_{1}(1400)}(Q^2)$}
      & -0.106 & 1.234 & 6.843&1.847
      \\
      \hline
  \mbox{$g^{(B)}_{B_s^*BK_{1}(1270)}(Q^2)$}
      & 0.764 & 0.412 & 11.343&4.708
      \\
       \hline
  \mbox{$g^{(B)}_{B_s^*BK_{1}(1400)}(Q^2)$}
      & 2.463 & -2.178 & 38.732&18.980
      \\
       \hline
  \mbox{$g^{(K_{1}(1270))}_{B_s^*BK_{1}(1270)}(Q^2)$}
      & 0.047 & -23681.595 & -31.416&2.914
      \\
      \hline
  \mbox{$g^{(K_{1}(1400))}_{B_s^*BK_{1}(1400)}(Q^2)$}
      & -0.021 & 0.282 & 3.080&1.233
      \\
    \hline \hline
\end{array}
$$
\caption{Parameters appearing in the fit function of the 
coupling constants for  $D_s^*DK_{1}(1270)$, $D_s^*DK_{1}(1400)$,
$B_s^*BK_{1}(1270)$ and $B_s^*BK_{1}(1400)$ vertices. $A_1$ and $A_2$ are in $GeV^{-1}$ units, while $x_0$ and $\Delta x$ are in the units of $GeV^2$.}
\label{fitparametrization}
\renewcommand{\arraystretch}{1}
\addtolength{\arraycolsep}{-1.0pt}
\end{table}
The same procedure as in the scalar and pseudoscalar cases leads to the numerical results for the corresponding coupling constants as presented in the Tables (\ref{gBsBK11400} and \ref{gBsBK1400}).
\begin{table}[h]
\renewcommand{\arraystretch}{1.5}
\addtolength{\arraycolsep}{3pt}
$$
\begin{array}{|c|c|c|c|}
\hline \hline
 \mbox{ } &
 Q^2=-m_D^2 &  Q^2=-m_{K_1(1270)}^2& \mbox{Average}

 \\
\hline
  \mbox{$g_{D_s^*DK_1(1270)}$}
      & 2.83\pm0.09 & 1.36\pm0.14 &
      2.09\pm0.82
      \\
    \hline \hline
    \mbox{ } &
 Q^2=-m_D^2 &  Q^2=-m_{K_1(1400)}^2& \mbox{Average}

 \\
 \hline
 \mbox{$g_{D_s^*DK_1(1400)}$}
      & 0.97\pm0.15 & 1.12\pm0.54 &
      1.04\pm0.78
      \\
    \hline \hline
\end{array}
$$
\caption{Values of the $g_{D_s^*DK_1(1270)}$ and
$g_{D_s^*DK_1(1400)}$ coupling constants in $GeV^{-1}$.} \label{gBsBK11400}
\renewcommand{\arraystretch}{1}
\addtolength{\arraycolsep}{-1.0pt}
\end{table}
\begin{table}[h]
\renewcommand{\arraystretch}{1.5}
\addtolength{\arraycolsep}{3pt}
$$
\begin{array}{|c|c|c|c|}
\hline \hline
 \mbox{ } &
 Q^2=-m_B^2 &  Q^2=-m_{K_1(1270)}^2&  \mbox{Average}

 \\
\hline
  \mbox{$g_{B_s^*BK_1(1270)}$}
      & 1.18\pm0.07 & 0.81\pm0.45 &
      1.99\pm0.11
      \\
    \hline \hline
    \mbox{ } &
 Q^2=-m_B^2 &  Q^2=-m_{K_1(1400)}^2&  \mbox{Average}

 \\
 \hline
 \mbox{$g_{B_s^*BK_1(1400)}$}
      & 0.35\pm0.05 & 0.26\pm0.04 &
      0.30\pm0.05
      \\
    \hline \hline
\end{array}
$$
\caption{Values of the $g_{B_s^*BK_1(1270)}$ and
$g_{B_s^*BK_1(1400)}$ coupling constants in $GeV^{-1}$.} \label{gBsBK1400}
\renewcommand{\arraystretch}{1}
\addtolength{\arraycolsep}{-1.0pt}
\end{table}
In summary, the strong coupling constants,  $g_{B_{s}BK_0^*}$, $g_{D_{s}DK_0^*}$,   $g_{B^{\ast}_{s}BK}$, $g_{D^{\ast}_{s}D K}$, $g_{B^{\ast}_{s}BK_1}$ and $g_{D^{\ast}_{s}D K_1}$,   have been calculated
 in the framework of three-point QCD sum rules. The
correlation functions of the considered vertices when  both $B(D)$ and $K_0^*(K)(K_1)$
mesons are off-shell are evaluated. The final numerical values have been obtained taking the average of the numerical values obtained from both off-shell cases. In the case of the axial vector $K_1$, 
which is either $K_1(1270)$ or $K_1(1400)$, the mixing between these two states have also been taken into account. A comparison of the obtained result on $D^{\ast}_{s}D K$  as the only previously calculated 
coupling 
constant among the considered strong coupling constants   has also been made.

\section{Acknowledgement}
 The authors thank E. Veli Veliev for his useful discussions. This work has been supported partly by the Scientific and Technological
Research Council of Turkey (TUBITAK) under the research project 110T284.

\end{document}